\documentclass[3p,times]{elsarticle}

\usepackage{ecrc}

\usepackage{epstopdf}

\volume{00}

\firstpage{1}

\journalname{Nuclear Physics A}

\runauth{D.Alexandre (for the ALICE Collaboration)}

\jid{nupha}

\jnltitlelogo{Nuclear Physics A}

\usepackage{graphicx}
\usepackage{amsmath,amssymb}
\usepackage{lineno}

\begin{document}

\begin{frontmatter}



\title{Multi-strange baryon production in pp, p--Pb and Pb--Pb
  collisions measured with ALICE at the LHC.}

\author{Didier Alexandre (for the ALICE\fnref{col1} Collaboration)}
\fntext[col1] {A list of members of the ALICE Collaboration and acknowledgements can be found at the end of this issue.}
\address{University of Birmingham, United Kingdom\\
E-mail: didier.alexandre@cern.ch}


\begin{abstract}
Multi-strange baryons are of particular interest in the understanding
of particle production mechanisms, as their high strangeness content
makes them susceptible to changes in the hadrochemistry of the
colliding systems. In ALICE, these hyperons are reconstructed via the
detection of their weak decay products, which are identified through their
measured ionisation losses and momenta in the Time Projection
Chamber.  The production rates of charged $\Xi$ and $\Omega$ baryons
in proton-proton (pp), proton-lead (p--Pb) and lead-lead (Pb--Pb)
collisions are reported as a function of $p_{\mathrm{T}}$. A direct
comparison in the hyperon-to-pion ratios between the three collision
systems is made as a function of event charged-particle
multiplicity. The recently measured production rates in p--Pb
interactions reveal an enhancement with increasing event multiplicity,
consistent with a hierarchy dependent on the strangeness content of
the hyperons. These results are discussed in the context of chemical
equilibrium predictions, taking into account the extracted temperature
parameter from a thermal model fit to the hadron yields in Pb--Pb data.     \\
\end{abstract}


\end{frontmatter} 



\vspace{0.5cm}
\section{Introduction}
\label{intro}

In pp collisions, the production of strange particles is
limited by the local canonical strangeness conservation law. This is
not the case in central heavy-ion collisions, where strange quarks are
expected to be produced more abundantly, which is translated to an
enhanced formation of strange baryons after hadronisation
\cite{BecattiniSHM, StrangeEnhance}. In addition, the
partial restoration of the chiral symmetry at the expected QCD phase
transition in A--A collisions lowers the energy threshold for strangeness
production to the level of the bare strange quark mass, a value which is
comparable to the temperatures achieved in such energetically dense
environments. For these reasons, the measurement of the charged
multi-strange $\Xi$ and $\Omega$ hyperons is an observable very
sensitive to the early stages of the collision system, and therefore
interesting for the understanding of strongly interacting matter in
different collision conditions.  \\

The phenomenon of strangeness enhancement in heavy-ion collisions with
respect to pp (p--Be) collisions has been observed with Pb--Pb interactions at
the NA57 experiment \cite{NA57}, at STAR \cite{STAR} with Au--Au collisions, and 
confirmed by ALICE in central 2.76 TeV Pb--Pb collisions
\cite{ALICE_multistrange_PbPb}. The chemical saturation values for the
$\Xi/\pi$ and $\Omega/\pi$ ratios\footnote{$\Xi/\pi$ and $\Omega/\pi$
  refer to the ratio of particle   and anti-particle sums.} extracted
from global fits to hadron yields in Au--Au collisions at STAR with
the grand canonical approach appeared to be consistent with the ALICE
central Pb--Pb results within systematic errors for an equilibrium
temperature of T=164 MeV \cite{PiKp_PbPb_2_Alice}. However, the same
model overpredicted the measured (p + $\bar{\mathrm{p}}$)/$(\pi^{-}+\pi^{+})$
values by about $50\%$. A lower chemical freeze-out
temperature of 156 MeV was obtained from the global fit to the
measured Pb--Pb hadron yields \cite{Stachel:2013zma}.  \\

ALICE has recently measured the multi-strange baryon production in the
intermediate p--Pb reference system as a function of charged-particle
multiplicity. In this conference proceeding, a description of the
method of identifying multi-strange baryons in p--Pb
collisions is given, the $p_{\mathrm{T}}$ spectra of these
particles are presented and the hyperon-to-pion
integrated yield ratios are explored as
a function of multiplicity in the three collision types. In addition,
these ratios at high multiplicity are compared to the chemical
equilibrium predictions with the grand canonical assumption.  \\

\section{Detection of multi-strange particles in p--Pb collisions with ALICE}
\label{techniques}

With an overall branching ratio of 63.9$\%$ (43.3$\%$), $\Xi^{-}$
($\Omega^{-}$) baryons weakly decay into a $\pi^{-}$ (K$^{-}$) and a
$\Lambda$, which itself decays into a p$\pi^{-}$ pair.\footnote{In
  the rest of this document, $\Xi^{-}$ and $\Omega^{-}$ will
  implicitly refer to both particle and antiparticle.} The tracks
formed by the daughter particles of these decays, are reconstructed in
the Inner Tracking System and in the large Time Projection Chamber
(TPC), and the particle identity is determined through the
ionisation losses and momenta measured in the TPC. \\

To reduce the combinatorial background in the selection of the weakly
decaying multi-strange particles ('cascades'), a set of geometrical
cuts based on the decay topologies were applied: minimum transverse
radial distances of the cascade and corresponding $\Lambda$ ('V0') decay
vertices form the Primary Vertex (PV) were set as part of the
selection criteria, as well as minimal Distances of Closest Approach
(DCA) between each final state particle track and the PV. Upper limit cuts
were set on the DCA between the V0 decay tracks and between the V0 and
the bachelor track. In addition, a cut on the cosine of pointing angle
- angle between the line connecting the PV to the decay vertex and the
reconstructed momentum vector of the decaying particle - of the 
$\Xi^{-}$ and of the daughter $\Lambda$ of $>0.97$ was
applied. \\ 

The signal was extracted from the invariant mass distributions
produced with the selected cascade candidates. In
every $p_{\mathrm{T}}$ bin, a Gaussian function was fit
to the mass peak and a central aera defined within $-4\sigma$
and $4\sigma$ of that peak. The candidate counts in two background bands defined
in the intervals [-8;-4]$\sigma$ and [4;8]$\sigma$ from the Gaussian peak were
subtracted from the entries in the central bins to give the signal. \\

The $p_{\mathrm{T}}$ spectra of the multi-strange particles were
measured in seven different multiplicity classes. The multiplicity
percentile of each event was obtained from the amplitude
distribution of the forward V0A detector (placed within the
pseudorapidity range $2.8<\eta<5.1$, on the outgoing Pb-beam
side). This avoided measuring the multiplicity in the same
mid-rapidity region of the detector as the region where the
multi-strange particles were measured (within the rapidity range
$-0.5<y<0.0$ in the centre-of-mass system of the collision system, and
each daughter track within $|\eta_{lab}|<0.8$), and possible selection
biases it could cause. The averaged charged-particle pseudorapidity
densities with respect to which the integrated multi-strange yields
were studied, were measured in the central $|\eta_{lab}|<0.5$ detector
region in each of the V0A -determined multiplicity classes.   \\

\section{Results}
\label{results}

\begin{figure}[h!]
\begin{center}
\includegraphics*[width=8.0cm,height=6.5cm]{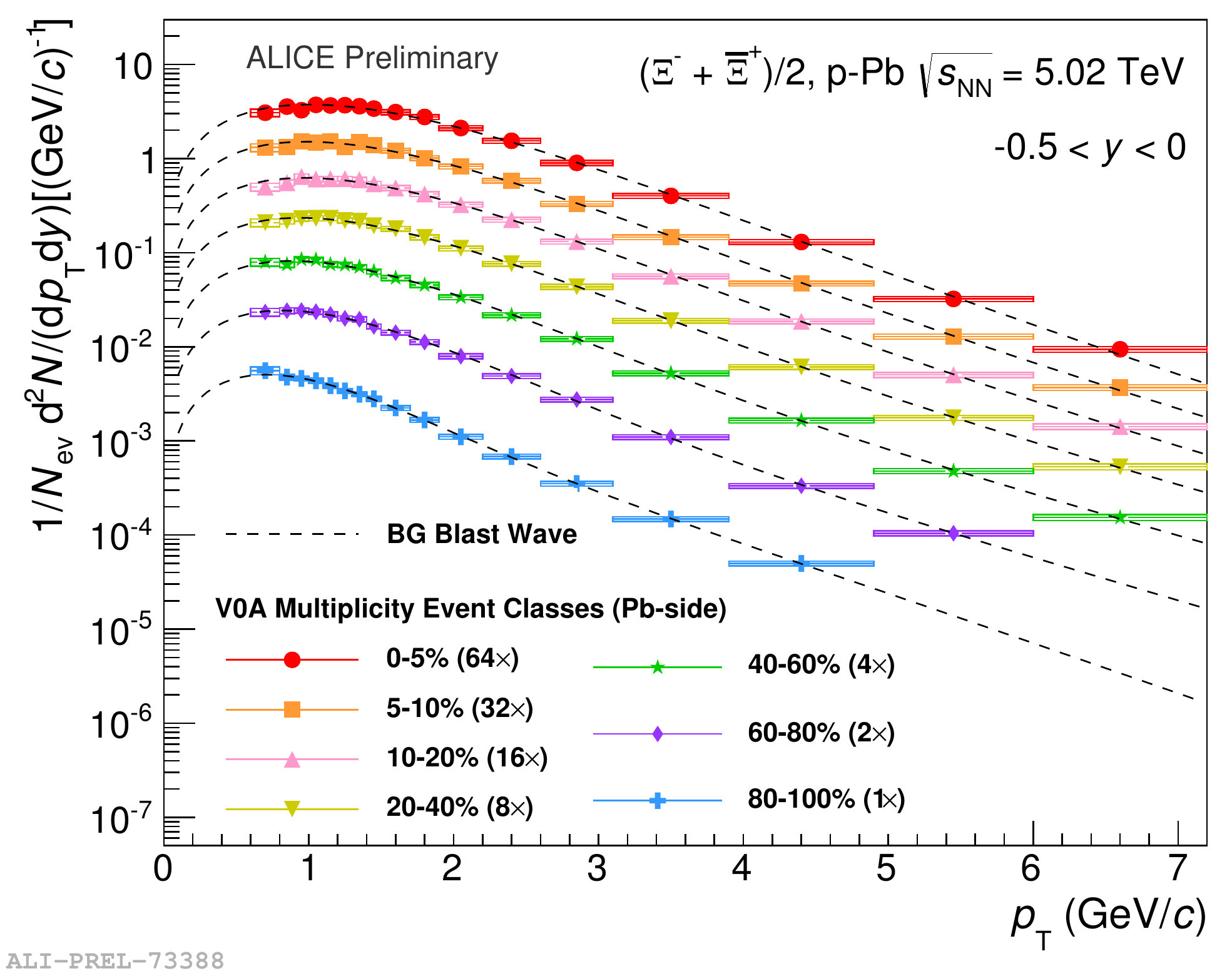}
\hspace{2mm}
\includegraphics*[width=8.0cm, height=6.5cm]{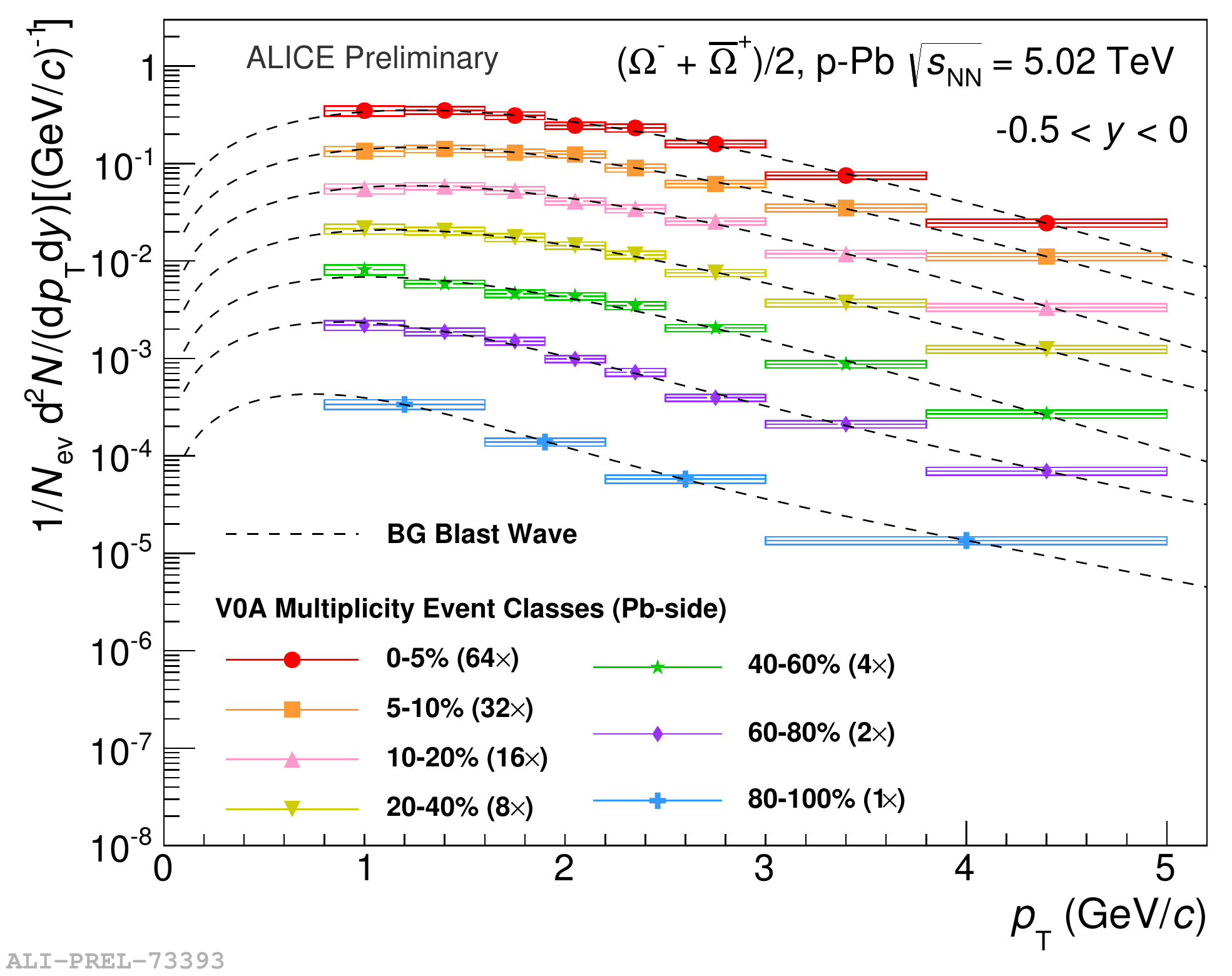}
\caption{Charged $\Xi$ and $\Omega$ $p_{\mathrm{T}}$ spectra in seven
  multiplicity classes. The average spectra of particle and
  anti-particle are shown.} 
\label{fig:Spectra}
\end{center}
\end{figure}

The measured $p_{\mathrm{T}}$ distributions ranging between 0.6 (0.8) and
7.2 (5.0) GeV/$\it{c}$ for $\Xi^{-}$ ($\Omega^{-}$) are shown in Figure
\ref{fig:Spectra} as a function of multiplicity. A hardening of the
spectra with multiplicity can be observed in the shape of the curves both for
$\Xi^{-}$ and $\Omega^{-}$. Moreover, the central $\Xi^{-}$ and
$\Omega^{-}$ p--Pb spectra have been shown to fit the
hydrodynamics-based Blast Wave (BW) function \cite{BW} simultaneously with the
other hadrons, yielding kinetic freeze-out temperature and mean expansion
velocity $<\beta_{T}>$ parameters in agreement with those already
published without the multi-strange particles
\cite{PiKpLambda_pPb_ALICE}. This hints at a collective behaviour of
the hyperons after the collision, consistent with the radial flow
interpretation from the p, $\pi$, K and $\Lambda$ production study in
p--Pb \cite{ALICE_multistrange_PbPb} and in Pb--Pb collisions
\cite{PiKp_PbPb_Alice,  K0SLambda_PbPb_ALICE}. Therefore a full-range
BW fit was used to extract the integrated yields and extrapolate the
$p_{\mathrm{T}}$ distributions down to 0 GeV/$\it{c}$. While the
$\Xi^{-}$ ($\Omega^{-}$) spectrum extrapolation fractions of the yields
amount to about 25$\%$ (40$\%$) in the lowest multiplicity classes
(80-100$\%$ percentile bands), the obtained errors due to the
extrapolation technique are of no more than 5$\%$ (15$\%$) in those
bins.     \\

\begin{figure}[h!]
\begin{center}
\includegraphics*[width=8.0cm, height=6.5cm]{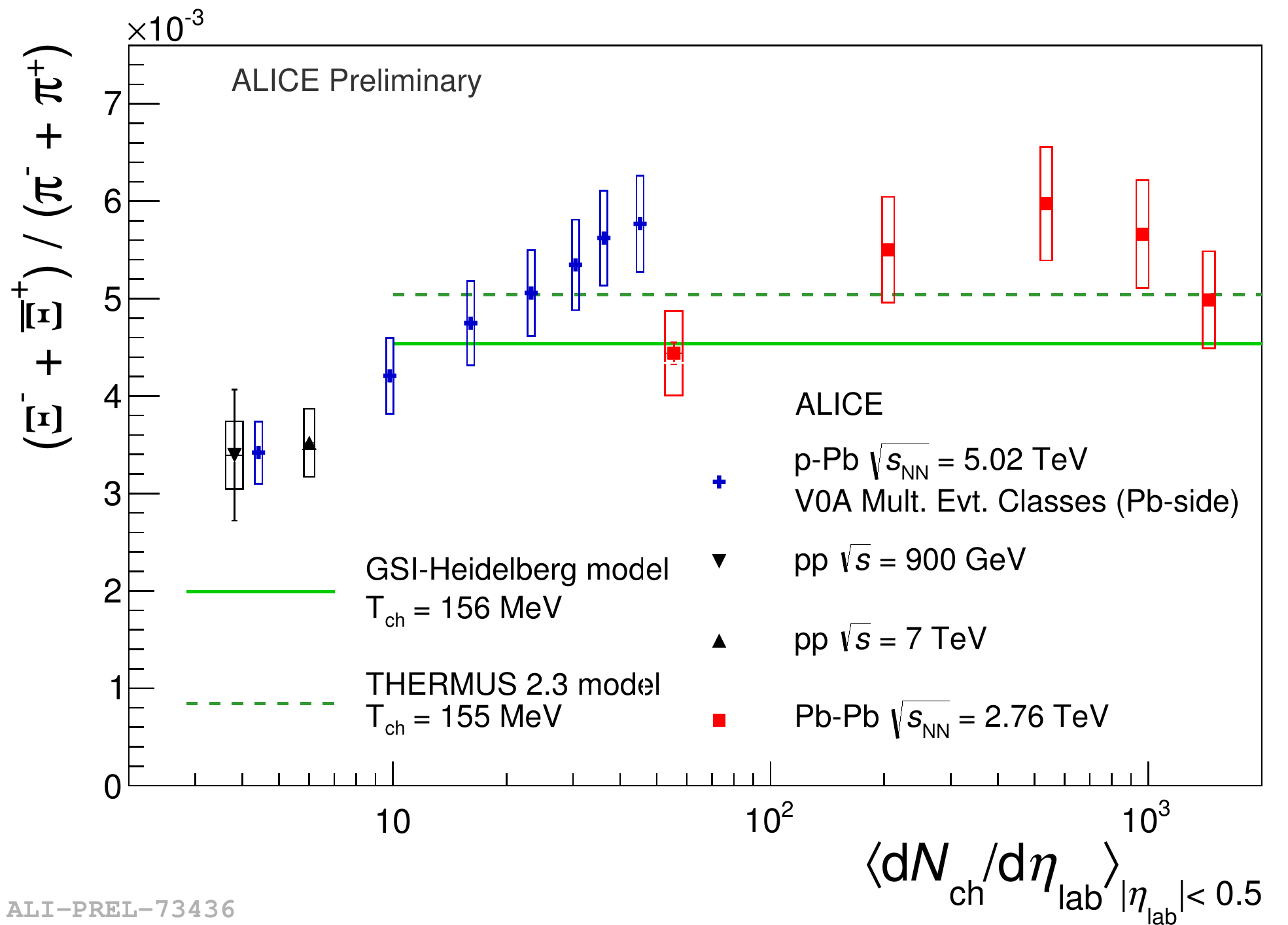}
\hspace{2mm}
\includegraphics*[width=8.0cm,
height=6.5cm]{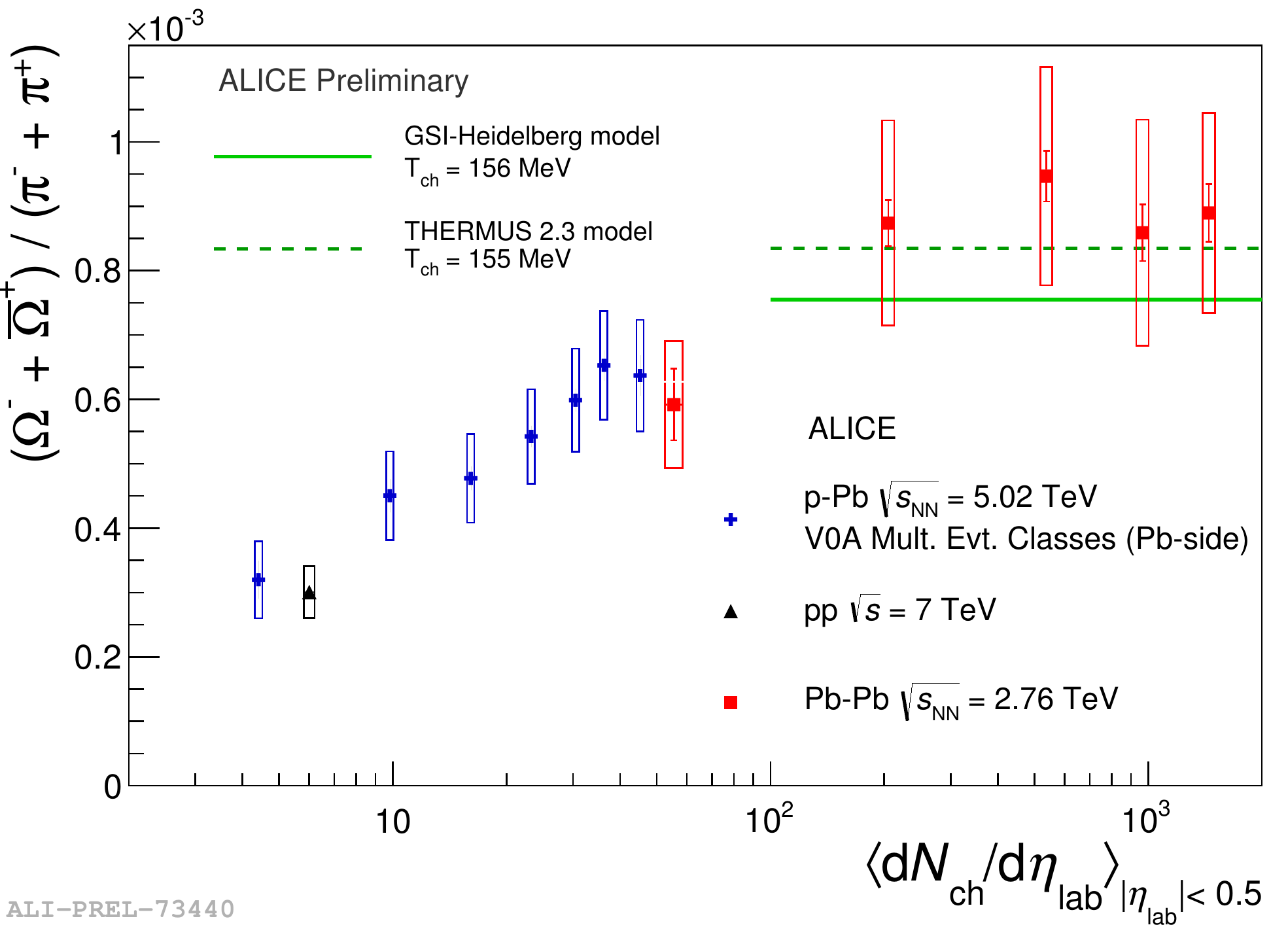}
\caption{($\Xi^{-}+\bar{\Xi}^{+}$)/($\pi^{-}+\pi^{+}$) and
  ($\Omega^{-}+\bar{\Omega}^{+}$)/($\pi^{-}+\pi^{+}$) ratios as a function
of charged-particle pseudorapidity density in pp, p--Pb and Pb--Pb collisions
measured by ALICE.}
\label{fig:StrEnhance}
\end{center}
\end{figure}

Figure \ref{fig:StrEnhance} shows the recently measured p--Pb hyperon-to-pion ratios
as a function of $\langle dN_{ch}/$d$\eta \rangle_{|\eta_{lab}|<0.5}$, compared with
pp and Pb--Pb measurements. The p--Pb data were taken in a
multiplicity range of over one order of magnitude and intermediate
between the other two collision systems. An enhancement of the
multi-strange baryons with multiplicity in p--Pb data is observed, the
effect being stronger for $\Omega$ baryons, for which the enhancement
factor in high with respect to low multiplicity events is roughly
2. In fact, together with the earlier $2\Lambda/(\pi^{-}+\pi^{+})$ ratios published
in \cite{PiKpLambda_pPb_ALICE}, one observes a consistent enhancement hierarchy in p--Pb
dependent on the strangeness quantum number of the baryon.  \\

The solid and dashed green lines in Figure \ref{fig:StrEnhance}
represent the grand canonical limits for chemical saturation resulting
from the global fits to ALICE Pb--Pb data. Two
different model limits are shown - the GSI-Heidelberg model \cite{GSImodel}, and the THERMUS
model \cite{ThermusModel} - with extracted chemical equilibrium temperatures of 156$\pm$1.5
MeV and 155$\pm$2.0 MeV, respectively \cite{MicheleFloris}. Except for the most peripheral
Pb--Pb data point, the saturation limits tend to overlap with the lower
end of the error bars of the Pb--Pb data points or lie below them. In p--Pb,
the $\Xi^{}/\pi^{}$ ratios in high multiplicity events agree with
those observed in non-peripheral Pb--Pb measurements. This stands in contrast
to the Omega-to-pion ratios, which lie below the values observed in
the most central Pb--Pb collisions. Consequently, the
$\Omega^{}/\pi^{}$ ratio does not reach the equilibrium level
suggested by the data.\\

\section{Conclusions}

The production of the multi-strange $\Xi^{-}$ and $\Omega^{-}$
baryons and their antiparticles was measured in p--Pb collisions with
the ALICE detector in seven different event multiplicity bins. A
hardening in the shape of the extracted p$_{T}$ distributions with increasing
multiplicity was observed, consistent with the interpretation of
increasing radial flow velocity. A clear strangeness enhancement with
respect to the pion production was measured in p--Pb collisions for these hadrons as a
function of charged-particle multiplicity. The p--Pb data points
bridge the ratio values in pp to those in Pb--Pb. A
theoretical description of this transition will improve our
understanding of strangeness production in high energy
collisions. The chemical equilibrium temperature parameter obtained
from the simultaneous fit to the hadron yields measured in ALICE Pb--Pb
collisions is lower than the one extracted at STAR with Au--Au data
and lies at about 156 MeV. While the $\Omega/\pi$ ratio in p--Pb
approaches this saturation level with increasing multiplicity, the high
multiplicity $\Xi/\pi$ values are comparable to those in central
Pb--Pb data and lie above the estimated saturation limits. \\ \\










\bibliographystyle{elsarticle-num}
\bibliography{references}


\end{document}